\documentclass{ws-mpla}

\begin{document}

\markboth{}
{}

%%%%%%%%%%%%%%%%%%%%% Publisher's Area please ignore %%%%%%%%%%%%%%
\catchline{}{}{}{}{}
%%%%%%%%%%%%%%%%%%%%%%%%%%%%%%%%%%%%%%%%%%%%%%%%%%%%%%%%%%%%%%%%%%%

\title{One loop correction to $Z\rightarrow \nu\nu$ in the Minimal R-symmetric Supersymmetric Standard Model}

\author{Ke-Sheng Sun    $^{a,\ast}$ ,
        Jian-Bin Chen   $^{b,\star}$,
        Xiu-Yi Yang     $^{c,\dagger}$,\\
        Hai-Bin Zhang   $^{d,\ddagger}$}

\address{$^{a}$   Department of Physics, Baoding University, Baoding 071000,China\\
         $^{b}$   College of Physics and Optoelectronic Engineering, Taiyuan University of Technology, Taiyuan 030024, China\\
         $^{c}$   School of Science, University of Science and Technology Liaoning, Anshan 114051, China\\
         $^{d}$   Department of Physics, Hebei University, Baoding 071002, China\\
         $^{\ast}   $    sunkesheng@126.com;sunkesheng@mail.dlut.edu.cn\\
         $^{\star}  $    chenjianbin@tyut.edu.cn\\
         $^{\dagger}$    yxyruxi@163.com\\
         $^{\ddagger}$   hbzhang@hbu.edu.cn}

\maketitle

\pub{Received (Day Month Year)}{Revised (Day Month Year)}

\begin{abstract}
We analyze the one loop correction to $Z\rightarrow \nu\nu$ decay in framework of Minimal R-symmetric Supersymmetric Standard Model(MRSSM) in detail with normal and inverse neutrino mass orderings, as a function of $\tan\beta$, Dirac mass parameters $M_D^W$ and $\mu_u(\mu_d)$, slepton mass $m_l$ that parameterize the mass matrices. The numerical results indicate that the branching ratio for $Z\rightarrow \nu\nu$ decay is compatible with the experimental measurement and the SM expectation at $2\sigma$ level. For inverse neutrino mass ordering, the prediction exceeds the SM expectation at $1\sigma$ level. The prediction on $Br(Z\rightarrow \nu\nu)$ increases proportionally to $\tan\beta$ and inversely proportionally to $m_l$. For normal neutrino mass ordering, the peak value of the prediction on $Br(Z\rightarrow \nu\nu)$ exceeds the SM expectation at $1\sigma$ level.
\end{abstract}

\section{Introduction}\label{intro}	

Invisible decay of Z boson offers constraints on models that lie beyond the Standard Model (SM).
The experiment measurement and the SM expectation for the branching ratio of invisible Z-width are \cite{PDG}
\begin{eqnarray}
BR(Z\rightarrow invisible)_{Exp} &=& 0.2000\pm 0.0006,\\
BR(Z\rightarrow invisible)_{SM} &=& 0.2010\pm 0.0001.
\end{eqnarray}
It shows that the current value of the invisible Z-width in experiment agrees quite well with the SM expectation within $2\sigma$ errors. This is often interpreted as evidence that the SM contains three and only three neutrinos cause, apart from neutrinos, the SM predicts no other channels that these states can decay into. However, there is still a $1\sigma$ level deviation between the experimental measurement and the SM expectation. This departure may hint structures over and above those in the SM.

As a new solution to the supersymmetric flavor problem in MSSM, the Minimal R-symmetric Supersymmetric Standard Model (MRSSM) is proposed in Ref.\cite{Kribs}, where the R-symmetry is a fundamental symmetry proposed several decades ago and stronger than R-parity \cite{Fayet,Salam}. R-symmetry forbids Majorana gaugino masses, $\mu$ term, $A$ terms and all left-right squark and slepton mass mixings. The $R$-charged Higgs $SU(2)_L$ doublets $\hat{R}_u$ and $\hat{R}_d$ are introduced in MRSSM to yield the Dirac mass terms of higgsinos.	Additional superfields $\hat{S}$, $\hat{T}$ and $\hat{O}$ are introduced to yield Dirac mass terms of gauginos. Studies on phenomenology in MRSSM can be found in literatures \cite{Die1,Die2,Die3,Die4,Die5,Kumar,Blechman,Kribs1,Frugiuele,Jan,Chakraborty,Braathen,Athron,Alvarado}.

In SM, the $Z\rightarrow \nu_i\nu_j(i=j)$ decays occur at tree level through the interaction between three neutrinos and $Z$ boson
\begin{eqnarray}
{\cal L}& = &C^{Tree}\sum_{i=e,\mu,\tau}\overline{\nu_{i}}\gamma_\mu P_L \nu_{j}Z^\mu,\nonumber\\
C^{Tree}&=&-\frac{1}{2}\delta^{ij}(g_1 s_W+g_2 c_W),\nonumber
\end{eqnarray}
where $s_W=sin\theta_{W}$, $c_W=cos\theta_{W}$ and $\theta_{W}$ is the Weinberg angle, $g_1$ denotes the coupling constant of gauge group U(1), $g_2$ denotes the coupling constant of gauge group SU(2). At one loop level, the $Z\rightarrow \nu_i\nu_j(i\ne j)$ decays arise from the charged current with the mixing among three lepton generations. The fields of the flavor neutrinos in charged current weak interaction Lagrangian are combinations of three massive neutrinos
\begin{eqnarray}
{\cal L}& = &-\frac{g_2}{\sqrt{2}}\sum_{l=e,\mu,\tau}\overline{l_{L}}(x)\gamma_\mu \nu_{lL}(x)W^\mu(x)+h.c.,
\nonumber\\
\nu_{lL}(x)& = &\sum_{i=1}^{3}\Big(U_{PMN}\Big)_{li}\nu_{iL}(x),\nonumber
\end{eqnarray}
where $\nu_{lL}$ are fields of the flavor neutrinos, $\nu_{iL}$ are fields of massive neutrinos, and $U_{PMN}$ corresponds to the unitary neutrino mixing matrix \cite{Pontecorvo,Maki}. The unitary neutrino mixing matrix is given by
\begin{eqnarray}
U_{PMN}
&=&\left(\begin{array}{ccc}
c_{1}c_{3}&c_{3}s_{1}&s_{3}e^{-i\delta}\\
-c_{1}s_{3}s_{2}e^{i\delta}-c_{2}s_{1}&c_{1}c_{2}-s_{1}s_{2}s_{3}e^{i\delta}&c_{3}s_{2}\\
s_{1}s_{2}-c_{1}s_{3}c_{2}e^{i\delta}&c_{1}s_{2}-s_{1}c_{2}s_{1}e^{i\delta}&c_{3}c_{2}
\end{array}\right)\nonumber\\
&&\times   diag\Big(e^{i\Phi_{1}/2},1,e^{i\Phi_{2}/2}\Big),\nonumber
\label{PMN}
\end{eqnarray}
where $s(c)_{1}$ = $\sin(\cos)\theta_{12}$, $s(c)_{2}$ = $\sin(\cos)\theta_{23}$, $s(c)_{3}$ = $\sin(\cos)\theta_{13}$. The phase $\delta$ is the Dirac phase, and $\Phi_{i}$ are the Majorana phases.

In this paper, we have studied the $Z\rightarrow \nu\nu$ in MRSSM, where the neutrino mass is assumed with normal ordering and inverse ordering. On one hand, similar to the case in MSSM, the $Z\rightarrow \nu\nu$ decays can arise from the diagrams mediated by charginos $\chi^{\pm}$, neutrinolinos $\chi^0$, sleptons $\tilde{L}^{\pm}$ and sneutrinos $\tilde{\nu}$. It is noted worthwhile that the mass matrices in MSSM are different from MRSSM. On the other hand, in MRSSM, the $Z\rightarrow \nu\nu$ decays can also arise from diagrams mediated by another two charginos $\rho^{\pm}$ which have R-charge minus electric charge. Assuming the parameter spaces with two set benchmark points, we investigate the $Z\rightarrow \nu\nu$ decays as functions of $\tan\beta$, Dirac mass parameters $M_D^W$, $\mu_u$, $\mu_d$ and slepton mass parameter $m_l$ with normal and inverse neutrino mass orderings. The result agrees with the experimental measurement and the SM expectation very well.

The paper is organized as follows. In Section \ref{sec2}, we provide a brief introduction on MRSSM, and derive the analytic expressions for every Feynman diagram contributing to $Z\rightarrow \nu_i\nu_j$ in MRSSM in detail. The numerical results are presented in Section \ref{sec3}, and the conclusion is drawn in Section \ref{sec4}.
\section{MRSSM}
\label{sec2}

The general form of the superpotential of the MRSSM is given by \cite{Die1}
\begin{eqnarray}
\mathcal{W}_{MRSSM} &=& \mu_d(\hat{R}_dH_d)+\mu_u(\hat{R}_uH_u)+\Lambda_d(\hat{R}_d\hat{T})H_d+\Lambda_u(\hat{R}_u\hat{T})H_u\nonumber\\
&+&Y_u\bar{U}(QH_u)-Y_d\bar{D}(QH_d)-Y_e\bar{E}(LH_d)\nonumber\\
&+&\lambda_d\hat{S}(\hat{R}_dH_d)+\lambda_u\hat{S}(\hat{R}_uH_u),
\end{eqnarray}
where $H_u$ and $H_d$ are the MSSM-like Higgs weak iso-doublets, $\hat{R}_u$ and $\hat{R}_d$ are the $R$-charged Higgs $SU(2)_L$ doublets and the corresponding Dirac higgsino mass parameters are denoted as $\mu_u$ and $\mu_d$. $\lambda_u$, $\lambda_d$, $\Lambda_u$ and $\Lambda_d$ are parameters of Yukawa-like trilinear terms involving the singlet $\hat{S}$ and the triplet $\hat{T}$, which is given by
\begin{equation}
\hat{T} = \left(
\begin{array}{cc}
\hat{T}^0/\sqrt{2} &\hat{T}^+ \\
\hat{T}^-  &-\hat{T}^0/\sqrt{2}\end{array}
\right).\nonumber
 \end{equation}
The soft-breaking scalar mass terms are given by
\begin{eqnarray}
V_{SB,S} &=& m^2_{H_d}(|H^0_d|^2+|H^{-}_d|^2)+m^2_{H_u}(|H^0_u|^2+|H^{+}_u|^2)+m^2_{R_u}(|R^0_u|^2+|R^{-}_u|^2)\nonumber\\
&+&m^2_{R_d}(|R^0_d|^2+|R^{+}_d|^2)+(B_{\mu}(H^-_dH^+_u-H^0_dH^0_u)+h.c.)\nonumber\\
&+&m^2_S|S|^2+ m^2_O|O^2|+m^2_T(|T^0|^2+|T^-|^2+|T^+|^2)\nonumber\\
&+&\tilde{d}^*_{L,i} m_{q,{i j}}^{2} \tilde{d}_{L,j} +\tilde{d}^*_{R,i} m_{d,{i j}}^{2} \tilde{d}_{R,j}+\tilde{u}^*_{L,i}  m_{q,{i j}}^{2} \tilde{u}_{L,j}+\tilde{u}^*_{R,i}  m_{u,{i j}}^{2} \tilde{u}_{R,j}\nonumber\\
&+&\tilde{e}^*_{L,i} m_{l,{i j}}^{2} \tilde{e}_{L,j}+\tilde{e}^*_{R,{i}} m_{r,{i j}}^{2} \tilde{e}_{R,{j}} +\tilde{\nu}^*_{L,i} m_{l,{i j}}^{2} \tilde{\nu}_{L,j}
\end{eqnarray}
All trilinear scalar couplings involving Higgs bosons to squarks and sleptons are forbidden due to the $R$-symmetry. The soft-breaking Dirac mass terms of the singlet $\hat{S}$, triplet $\hat{T}$ and octet $\hat{O}$ take the form
\begin{equation}
V_{SB,DG}=M^B_D\tilde{B}\tilde{S}+M^W_D\tilde{W}^a\tilde{T}^a+M^O_D\tilde{g}\tilde{O}+h.c.,
\label{}
\end{equation}
where $\tilde{B}$, $\tilde{W}$ and $\tilde{g}$ are usually MSSM Weyl fermions. After EWSB, the mass matrix of four neutralinos $\chi^0_{1,2,3,4}$ is given by
\begin{eqnarray}
m_{\chi^0} &=& \left(
\begin{array}{cccc}
M^{B}_D &0 &-\frac{1}{2} g_1 v_d  &\frac{1}{2} g_1 v_u \\
0 &M^{W}_D &\frac{1}{2} g_2 v_d  &-\frac{1}{2} g_2 v_u \\
- \frac{1}{\sqrt{2}} \lambda_d v_d  &-\frac{1}{2} \Lambda_d v_d  &-\mu_d^{eff,+}&0\\
\frac{1}{\sqrt{2}} \lambda_u v_u  &-\frac{1}{2} \Lambda_u v_u  &0 &\mu_u^{eff,-}\end{array}
\right),
 \end{eqnarray}
where the modified $\mu_i$ parameters are given by
\begin{align}
\mu_d^{eff,+}&= \frac{1}{2} \Lambda_d v_T  + \frac{1}{\sqrt{2}} \lambda_d v_S  + \mu_d ,\nonumber\\
\mu_u^{eff,-}&= -\frac{1}{2} \Lambda_u v_T  + \frac{1}{\sqrt{2}} \lambda_u v_S  + \mu_u.\nonumber
\end{align}
The $v_T$ and $v_S$ are vacuum expectation values of $\hat{T}$ and $\hat{S}$ which carry zero $R$-charge.
The neutralino mass matrix is diagonalized by unitary matrices $N^1$ and $N^2$
\begin{equation}
(N^{1})^{\ast} m_{\chi^0} (N^{2})^{\dagger} = diag(m_{\chi^0_1},...,m_{\chi^0_4}).\nonumber
\end{equation}
The mass matrix of two charginos $\chi^{\pm}_{1,2}$ with R-charge equal to electric charge is given by
\begin{equation}
m_{\chi^{\pm}} = \left(
\begin{array}{cc}
g_2 v_T  + M^{W}_D &\frac{1}{\sqrt{2}} \Lambda_d v_d \\
\frac{1}{\sqrt{2}} g_2 v_d  &\mu_u^{eff,-}\end{array}
\right),
 \end{equation}
and can be diagonalized by unitary matrices $U^1$ and $V^1$
\begin{equation}
(U^{1})^{\ast} m_{\chi^{\pm}} (V^{1})^{\dagger} =diag(m_{\chi^{\pm}_1},m_{\chi^{\pm}_2}).\nonumber
\end{equation}
The mass matrix of two charginos $\rho^{\pm}_{1,2}$ with R-charge equal to minus electric charge is given by
\begin{equation}
m_{\rho^{\pm}} = \left(
\begin{array}{cc}
-g_2 v_T  + M^{W}_D &\frac{1}{\sqrt{2}} g_2 v_u \\
-\frac{1}{\sqrt{2}} \Lambda_u v_u   &-\mu_u^{eff,+}\end{array}
\right),
 \end{equation}
and can be diagonalized by unitary matrices $U^2$ and $V^2$
\begin{equation}
(U^{2})^{\ast} m_{\rho^{\pm}} (V^{2})^{\dagger} =diag(m_{\rho^{\pm}_1},m_{\rho^{\pm}_2}).\nonumber
\end{equation}
In the gauge eigenstate basis $\tilde{\nu}_{iL}$, the sneutrino mass squared matrix is expressed as
\begin{equation}
m^2_{\tilde{\nu}} =
\begin{array}{c}m_l^2+\frac{1}{8}(g_1^2+g_2^2)( v_{d}^{2}- v_{u}^{2})+g_2 v_T M^{W}_D-g_1 v_S M^{B}_D,
\end{array}
 \end{equation}
where the last two terms are newly introduced by MRSSM, and is diagonalized by unitary matrix $Z^V$
\begin{equation}
Z^V m^2_{\tilde{\nu}} (Z^{V})^{\dagger} = diag(m^{2}_{\tilde{\nu}_1},m^{2}_{\tilde{\nu}_2},m^{2}_{\tilde{\nu}_3}).\nonumber
\end{equation}
The slepton mass squared matrix is given by
\begin{equation}
m^2_{\tilde{L}^{\pm}} = \left(
\begin{array}{cc}
(m^2_{\tilde{L}^{\pm}})_{LL} &0 \\
0  &(m^2_{\tilde{L}^{\pm}})_{RR}\end{array}
\right),
 \end{equation}
with
\begin{align}
(m^2_{\tilde{L}^{\pm}})_{LL} &=m_l^2+ \frac{1}{2} v_{d}^{2} |Y_{e}|^2 +\frac{1}{8}(g_1^2-g_2^2)(v_{d}^{2}- v_{u}^{2}) -g_1 v_S M_D^B-g_2v_TM_D^W ,\nonumber\\
(m^2_{\tilde{L}^{\pm}})_{RR} &= m_r^2+\frac{v_d^2}{2}|Y_e|^2+\frac{1}{4}g_1^2( v_{u}^{2}- v_{d}^{2})+2g_1v_SM_D^B.\nonumber
\end{align}
One can see that the left-right slepton mass mixing is absent. The slepton mass matrix is diagonalized by unitary matrix $Z^E$
\begin{equation}
Z^E m^2_{\tilde{L}^{\pm}} (Z^{E})^{\dagger} =diag(m^2_{\tilde{L}^{\pm}_1},...,m^2_{\tilde{L}^{\pm}_6}). \nonumber
\end{equation}
%%%%%%%%%%%%%%%%%%%%%%%%%%%%%%%%%%%%%%%%%%%%%%%%%%%%%%%%%%%%%%%%%%%
\begin{figure}[htbp]
\setlength{\unitlength}{1mm}
\centering
\begin{minipage}[c]{1\textwidth}
\includegraphics[width=5.0in]{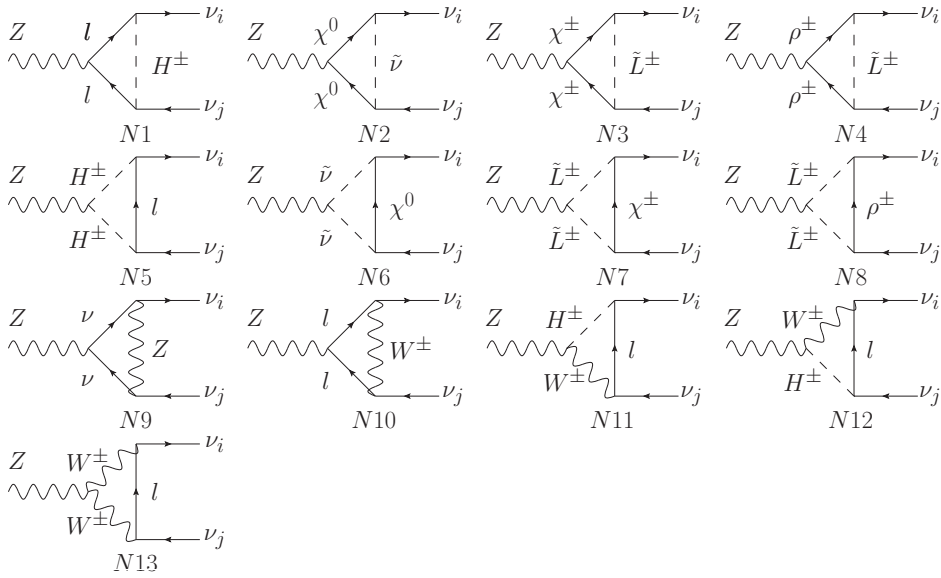}
\end{minipage}
\caption[]{One loop Feynman diagrams contributing to $Z\rightarrow \nu_i\nu_j$ in MRSSM.}
\label{Znunu}
\end{figure}
%%%%%%%%%%%%%%%%%%%%%%%%%%%%%%%%%%%%%%%%%%%%%%%%%%%%%%%%%%%%%%%%%%%

The relevant Feynman diagrams contributing to $Z\rightarrow \nu_i\nu_j$ in MRSSM is presented in Fig.\ref{Znunu}. The $Z \nu_i\nu_j$ interaction Lagrangian can be written as \cite{Flavor}
\begin{eqnarray}
\mathcal{L}_{Z \nu_i\nu_j}=\bar{\nu}_i\Big[\gamma^{\mu}(C^1_L P_L+ C^1_R P_R)+p_1^{\mu}(C^2_L P_L+ C^2_R P_R)\Big]\nu_jZ_{\mu}.
\end{eqnarray}
Then the branching ratio of LFV decays of $Z$ boson is calculated by
\begin{eqnarray}
Br(Z\rightarrow \nu_i\nu_j)&=&Br(Z\rightarrow \bar{\nu}_i \nu_j)+Br(Z\rightarrow \bar{\nu}_j \nu_i)\nonumber\\
&=&\frac{m_Z}{48\pi\Gamma_Z}\Big[2(|C^1_L|^2+|C^1_R|^2)+\frac{m_Z^2}{4}(|C^2_L|^2+|C^2_R|^2)\Big],
\end{eqnarray}
where the neutrino masses have been neglected and $\Gamma_Z$ is the total decay width of $Z$ boson. For convenience, following notation is used
\begin{eqnarray}
Br(Z\rightarrow \nu\nu)=\sum_{i,j=1,2,3}Br(Z\rightarrow \nu_i\nu_j),i\le j.\nonumber
\end{eqnarray}
The coefficients $C^{1}_{L/R}$ and $C^{2}_{L/R}$ are combinations of coefficients corresponding to each Feynman diagram in Fig.\ref{Znunu}
\begin{eqnarray}
C^{1}_{L}&=&C^{Tree}+\sum_{N=N1}^{N13}C^{1,N}_{L},\nonumber\\
C^{1}_{R}&=&\sum_{N=N1}^{N13}C^{1,N}_{R},\nonumber\\
C^{2}_{L/R}&=&\sum_{N=N1}^{N13}C^{2,N}_{L/R}.\nonumber
\end{eqnarray}
Actually, only $C^{1,N}_{L}$ contribute to the decay width cause other coefficients $C^{1,N}_{R}$ and $C^{2,N}_{L/R}$ equal zero. The explicit expressions of $C^{1,N}_{L}$ are derived by assuming the three neutrino masses are zero in terms of invariant Passarino-Veltman integrals \cite{PVI}. The coefficients $C^{1,N}_{L}$ in Fig.\ref{Znunu} (N1-N4) are calculated by
\begin{eqnarray}
 C^{1,N}_{L}=C^N_{1R}C^N_{3L}(C^N_{2R}\mathcal{B}_0(m_Z^2,M_1,M_2)+(C^N_{2L} M_1 M_2 +C^N_{2R} M_3^2)\mathcal{C}_0-2C^N_{2R}\mathcal{C}_{00}),\nonumber
\end{eqnarray}
where
\begin{eqnarray}
C^{N1}_{1R}&=&Y^i_lZ^+_{n1},C^{N1}_{2L}=\frac{\delta^{ij}}{2}(g_2 c_W-g_1 s_W),C^{N1}_{2R}=-g_1 s_W\delta^{ij}, \nonumber\\
C^{N1}_{3L}&=&Y^j_lZ^+_{n1},M_1=m_{l_k},M_2=m_{l_m},M_3=m_{H^{\pm}_{n}},(N=N1),\nonumber\\
C^{N2}_{1R}&=&\frac{1}{\sqrt{2}}Z^{V,\ast}_{ni}(g_1 N^1_{k1}-g_2N^1_{k2}),C^{N2}_{3L}=\frac{1}{\sqrt{2}}Z^{V}_{nj}(g_1 N^{1,\ast}_{m1}-g_2N^{1,\ast}_{m2}),\nonumber\\
C^{N2}_{2L}&=&\frac{1}{2}(g_2 c_W+g_1 s_W)(N^{2,\ast}_{k3}N^{2,\ast}_{m3}-N^{2,\ast}_{k4}N^{2,\ast}_{m4}), \nonumber\\
C^{N2}_{2R}&=&\frac{1}{2}(g_2 c_W+g_1 s_W)(N^{1,\ast}_{m3}N^{1,\ast}_{k3}-N^{1,\ast}_{m4}N^{2,\ast}_{k4}),\nonumber\\
M_1&=&m_{\chi^0_{k}},M_2=m_{\chi^0_{m}},M_3=m_{\tilde{\nu}_{n}},(N=N2),\nonumber
\end{eqnarray}
and
\begin{eqnarray}
C^{N3}_{1R}&=&Y^i_lZ^{E,\ast}_{n(3+i)}U^1_{k2},C^{N3}_{2L}=-\frac{1}{2}(2g_2c_WV^{1,\ast}_{m1}V^{1}_{k1}+(g_2 c_W-g_1s_W)V^{1,\ast}_{m2}V^{1}_{k2}),\nonumber\\
C^{N3}_{2R}&=&-\frac{1}{2}(2g_2c_WU^{1,\ast}_{k1}U^{1}_{m1}+(g_2 c_W-g_1s_W)U^{1,\ast}_{k2}U^{1}_{m2}),\nonumber\\
C^{N3}_{3L}&=&Y^j_lZ^{E}_{n(3+j)}U^{1,\ast}_{m2},M_1=m_{\chi^{\pm}_k},M_2=m_{\chi^{\pm}_m},M_3=m_{\tilde{L}^{\pm}_n},(N=N3),\nonumber\\
C^{N4}_{1R}&=&-g_2Z^{E,\ast}_{ni}U^2_{k1},C^{N4}_{2L}=\frac{1}{2}(2g_2c_WV^{2,\ast}_{m1}V^{2,\ast}_{k1}+(g_2 c_W-g_1 s_W)V^{2,\ast}_{m2}V^2_{k2}),\nonumber\\
C^{N4}_{2R}&=&\frac{1}{2}(2g_2c_WU^{2,\ast}_{k1}U^{2,\ast}_{m1}+(g_2 c_W-g_1 s_W)U^{2,\ast}_{k1}U^{2}_{m1}),C^{N4}_{3L}=-g_2U^{2,\ast}_{m1}Z^E_{nj},\nonumber\\
M_1&=&m_{\rho^{\pm}_{k}},M_2=m_{\rho^{\pm}_{m}},M_3=m_{\tilde{L}^{\pm}_n},(N=N4).\nonumber
\end{eqnarray}
The coefficients $C^{1,N}_{L}$ in Fig.\ref{Znunu} (N5-N8) are calculated by
\begin{eqnarray}
C^{1,N}_{L}=2C^N_{1R}C^N_2C^N_{3L}\mathcal{C}_{00},\nonumber
\end{eqnarray}
where
\begin{eqnarray}
C^{N5}_{1R}&=&Y^i_lZ^+_{k1},C^{N5}_{3L}=Y^j_lZ^+_{m1},\nonumber\\
C^{N5}_2&=&\frac{1}{2}((g_2 c_W-g_1s_W)(Z^+_{k1}Z^+_{m1}+Z^+_{k2}Z^+_{m2})+2g_2c_W(Z^+_{k3}Z^+_{m3}+Z^+_{k4}Z^+_{m4})),\nonumber\\
M_1&=&m_{H^{\pm}_{k}},M_2=m_{H^{\pm}_{m}},M_3=m_{l_n},(N=N5),\nonumber\\
C^{N6}_{1R}&=&\frac{1}{\sqrt{2}}Z^{V,\ast}_{ki}(g_1 N^1_{n1}-g_2N^1_{n2}),C^{N6}_{3L}=\frac{1}{\sqrt{2}}Z^{V}_{mj}(g_1 N^{1,\ast}_{n1}-g_2N^{1,\ast}_{n2}),\nonumber\\
C^{N6}_{2}&=&-\frac{1}{2}\delta^{km}(g_1 s_W+g_2 c_W),M_1=m_{\tilde{\nu}_k},M_2=m_{\tilde{\nu}_m},M_3=m_{\chi^{\pm}_n},(N=N6),\nonumber
\end{eqnarray}
and
\begin{eqnarray}
C^{N7}_{1R}&=&Y^i_lZ^{E,\ast}_{k(3+i)}U^1_{n2},C^{N7}_{3L}=Y^j_lZ^{E}_{m(3+j)}U^{1,\ast}_{n2},\nonumber\\
C^{N7}_2&=&\sum_{a=1,2,3}\frac{1}{2}(-2g_1s_WZ^{E,\ast}_{m(3+a)}Z^{E}_{k(3+a)}+(g_2 c_W-g_1s_W)Z^{E,\ast}_{ma}Z^{E}_{ka}),\nonumber\\
M_1&=&m_{\tilde{L}^{\pm}_k},M_2=m_{\tilde{L}^{\pm}_m},M_3=\chi^{\pm}_n,(N=N7),\nonumber\\
C^{N8}_{1R}&=&-g_2Z^{E,\ast}_{ki}U^2_{n1},C^{N8}_{3L}=-g_2Z^{E}_{mj}U^{2,\ast}_{n1},\nonumber\\
C^{N8}_2&=&\sum_{a=1,2,3}\frac{1}{2}(-2g_1s_WZ^{E,\ast}_{m(3+a)}Z^{E}_{k(3+a)}+(g_2 c_W-g_1s_W)Z^{E,\ast}_{ma}Z^{E}_{ka}),\nonumber\\
M_1&=&m_{\tilde{L}^{\pm}_k},M_2=m_{\tilde{L}^{\pm}_m},M_3=m_{\rho^{\pm}_n},(N=N8).\nonumber
\end{eqnarray}
The coefficients $C^{1,N}_{L}$ in Fig.\ref{Znunu} (N9-N10) are calculated by
\begin{eqnarray}
C^{1,N}_{L}&=&-2C^N_{1L}C^N_{3L}(-C^N_{2L}\mathcal{B}_0(m_Z^2,M_1,M_2)+C^N_{2L}(\mathcal{B}_0(0,M_3,M_1)+\mathcal{B}_0(0,M_3,M_2))\nonumber\\
&+&C^N_{2L}(M_1^2+M_2^2-M_3^2-m_Z^2)\mathcal{C}_0+C^N_{2R}M_1M_2\mathcal{C}_0-2C^N_{2L}\mathcal{C}_{00}),\nonumber
\end{eqnarray}
where
\begin{eqnarray}
C^{N9}_{1L}&=&-\frac{\delta^{ki}}{2}(g_1s_W+g_2c_W),C^{N9}_{2L}=-\frac{\delta^{km}}{2}(g_1s_W+g_2c_W),C^{N9}_{2R}=0,\nonumber\\
C^{N9}_{3L}&=&-\frac{\delta^{mj}}{2}(g_1s_W+g_2c_W),M_1=0,M_2=0,M_3=m_{Z},(N=N9),\nonumber\\
C^{N10}_{1L}&=&-\frac{ g_2}{\sqrt{2}} (U^{\ast}_{PMN})_{ki},C^{N10}_{2L}=\frac{\delta^{km}}{2}(g_2c_W-g_1s_W),C^{N10}_{2R}=-\delta^{km}g_1s_W,\nonumber\\
C^{N10}_{3L}&=&-\frac{g_2}{\sqrt{2}} (U_{PMN})_{mj},M_1=m_{l_k},M_2=m_{l_m},M_3=m_W,(N=N10).\nonumber
\end{eqnarray}
The coefficients $C^{1,N}_{L}$ in Fig.\ref{Znunu} (N11-N12) are calculated by
\begin{eqnarray}
C^{1,N}_{L}&=&C^N_1C^N_2C^N_3\mathcal{C}_{0},\nonumber
\end{eqnarray}
where
\begin{eqnarray}
C^{N11}_1&=&Y^i_lZ^+_{k1},C^{N11}_3=-\frac{g_2}{\sqrt{2}}(U_{PMN})_{nj},\nonumber\\
C^{N11}_2&=&-\frac{g_2 }{2}(-g_1 v_d s_W Z^+_{k1}+g_1 v_u s_W Z^+_{k2}+\sqrt{2}g_2 v_T c_W (Z^+_{k3}+Z^+_{k4}),\nonumber\\
M_1&=&m_{H^{\pm}_k},M_2=m_W,M_3=m_{l_n},(N=N11),\nonumber\\
C^{N12}_1&=&-\frac{ g_2}{\sqrt{2}}(U^{\ast}_{PMN})_{ki},C^{N12}_3=Y^j_lZ^+_{m1},\nonumber\\
C^{N12}_2&=&-\frac{ g_2}{2}(-g_1 v_d s_W Z^+_{m1}+g_1 v_u s_W Z^+_{m2}+\sqrt{2}g_2 v_T c_W (Z^+_{m3}+Z^+_{m4}),\nonumber\\
M_1&=&m_W,M_2=m_{H^{\pm}_m},M_3=m_{l_n},(N=N12).\nonumber
\end{eqnarray}
The coefficients $C^{1,N}_{L}$ in Fig.\ref{Znunu} (N13) are calculated by
\begin{eqnarray}
C^{1,N}_{L}&=&2C^N_1C^N_2C^N_3(\mathcal{B}_{0}(m_Z^2,M_1,M_2)+M_3^2\mathcal{C}_0+2\mathcal{C}_{00}),\nonumber
\end{eqnarray}
where
\begin{eqnarray}
C^{N13}_1&=&-\frac{ g_2}{\sqrt{2}}(U^{\ast}_{PMN})_{ni},C^{N13}_2=-g_2c_W,C^{N13}_3=-\frac{g_2 }{\sqrt{2}}(U_{PMN})_{nj},\nonumber\\
M_1&=&m_W,M_2=m_W,M_3=m_{l_n},(N=N13).\nonumber
\end{eqnarray}
The loop integrals are given in term of Passarino-Veltman \cite{PVI}
\begin{eqnarray}
\mathcal{C}_{(0,00)}&=&\frac{i}{16\pi^2}\mathcal{C}_{(0,00)}(0,m_Z^2,0;M_3,M_1,M_2).\nonumber
\end{eqnarray}
The explicit expressions of these loop integrals are given in Refs \cite{collier1,collier2,collier3} and $\overline{MS}$ scheme is used to delete the infinite terms. These loop integrals can be calculated through the Mathematica package Package-X \cite{X} and a link to Collier which is a fortran library for the numerical evaluation of one-loop scalar and tensor integrals\cite{collier}.
\section{Numerical Analysis}
\label{sec3}
In the numerical analysis, we use the two set benchmark points taken from existing references as the default values for our parameter setup and display them in Table.\ref{BMP} \cite{Die3}, where the slepton mass matrices are diagonal and all mass parameters are in $GeV$ or $GeV^2$. The following numerical values are used
\begin{eqnarray}
&&\alpha_{em}(m_Z)=1/128, m_Z=91.1876 GeV, m_W=80.379 GeV, \Gamma_{Z}=2.49 GeV, \nonumber\\
&&sin^2\theta_W=0.23129,m_e=0.510 MeV, m_{\mu}=105.6 MeV, m_{\tau}=1.776 GeV.\nonumber
\end{eqnarray}
The light neutrino mass spectrum is assumed to be normal ordering, the best-fit values of the three neutrino oscillation parameters are given in Table.\ref{BF}.
\begin{table}[h]
\tbl{The best-fit values correspond to normal ordering(NO) and inverse ordering(IO).}
{\begin{tabular}{@{}cccccc@{}}
\toprule
Parameter&best-fit(NO)&best-fit(IO)&Parameter&best-fit(NO)&best-fit(IO)\\
\colrule
$sin^2\theta_{12}$&0.297&0.297&$sin^2\theta_{23}$&0.425&0.589\\
$sin^2\theta_{13}$&0.0215&0.0216&$\delta$&1.38$\pi$&1.38$\pi$\\ \botrule
\end{tabular}\label{BF}}
\end{table}
Apart from the above result, no other experimental information on the Majorana phases in
the neutrino mixing matrix is available at present. Note that large value of $|v_T|$ is excluded by measurement of $W$ mass cause the vev $v_T$ of the $SU(2)_L$ triplet field $T^0$ gives a correction to $W$ mass through \cite{Die1}
\begin{eqnarray}
m_W^2=\frac{1}{4}g_2^2v^2+g_2^2v_T^2,
\label{}
\end{eqnarray}
with $v^2=v_u^2+v_d^2$.
\begin{table}[h]
\tbl{Benchmark points.}
{\begin{tabular}{@{}ccccccccccc@{}}
\toprule
Input&$tan\beta$&$\lambda_d$,$\lambda_u$&$\Lambda_d$,$\Lambda_u$&$v_S$&$v_T$&$M_D^B$&$M_D^W$&$\mu_d$,$\mu_u$&$m_T^2$&$m^2_l$,$m_r^2$\\
\colrule
BMP1&10&1.1,-1.1&-1.0,-1.0&1.3&-0.19&1000&500&400,400&$3000^2$&$1000^2$,$1000^2$\\
\colrule
BMP2&40&0.15,-0.15&-1.0,-1.15&-0.14&-0.34&250&500&400,400&$3000^2$&$1000^2$,$1000^2$\\
 \botrule
\end{tabular}\label{BMP}}
\end{table}
%%%%%%%%%%%%%%%%%%%%%%%%%%%%%%%%%%%%%%%%%%%%%%%%%%%%%%%%%%%%%%%%%%%
\begin{figure*}
\centering
\includegraphics[width=3.8in]{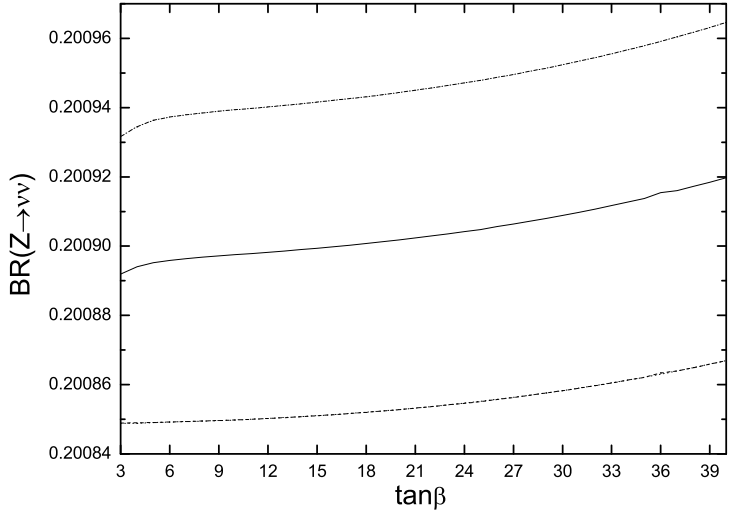}
\caption[]{The $Br(Z\rightarrow \nu\nu)$ vary as a function of $tan\beta$ in MRSSM, where the solid line (BMP1) and the dash dot line (BMP2) correspond to normal neutrino mass ordering, the dot line (BMP1) and the dash line (BMP2) correspond to inverse neutrino mass ordering and the two lines are almost identical. }
\label{fig1}
\end{figure*}
%%%%%%%%%%%%%%%%%%%%%%%%%%%%%%%%%%%%%%%%%%%%%%%%%%%%%%%%%%%%%%%%%%%

In Fig.\ref{fig1}, we display the theoretical prediction of $Br(Z\rightarrow \nu\nu)$ versus $tan\beta$ in MRSSM. The solid line and the dot line stand for the result calculated with parameter setup BMP1 with normal ordering and inverse ordering respectively. The dash dot line and the dash line stand for the result calculated with parameter setup BMP2 with normal ordering and inverse ordering respectively. Both predictions increase along with $tan\beta$ and are compatible with the experimental measurement at $2\sigma$ level. For normal ordering , the prediction of $Br(Z\rightarrow \nu\nu)$ with BMP1 is a little lower than that with BMP2. For inverse ordering, the predictions are almost identical. At $1\sigma$ level, the prediction of $Br(Z\rightarrow \nu\nu)$ with normal ordering is compatible with the SM expectation, however the prediction of $Br(Z\rightarrow \nu\nu)$ with inverse ordering exceed the SM expectation. There is a deviation between the predictions in MRSSM and the experimental measurement at $1\sigma$ level.

%%%%%%%%%%%%%%%%%%%%%%%%%%%%%%%%%%%%%%%%%%%%%%%%%%%%%%%%%%%%%%%%%%%
\begin{figure*}
\centering
\includegraphics[width=4.0in]{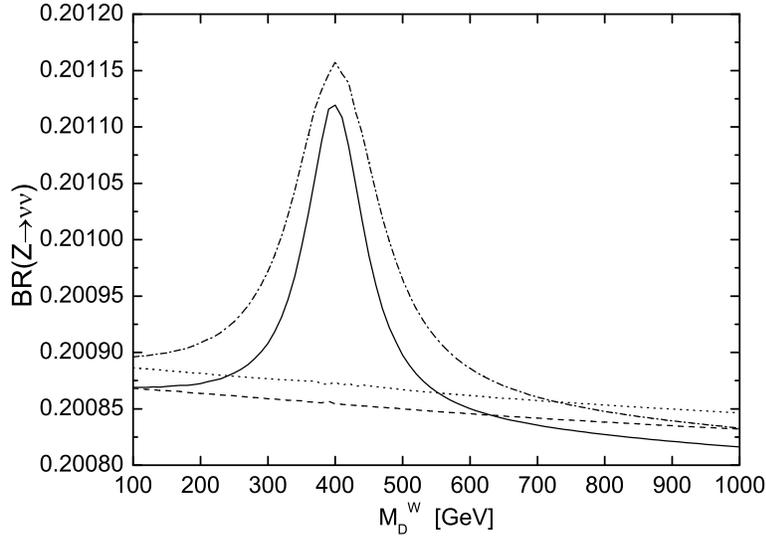}
\caption[]{The $Br(Z\rightarrow \nu\nu)$ vary as a function of $M_D^W$ in MRSSM, where the solid line (BMP1) and the dash dot line (BMP2) correspond to normal neutrino mass ordering, the dot line (BMP1) and the dash line (BMP2) correspond to inverse neutrino mass ordering. }
\label{fig2}
\end{figure*}
%%%%%%%%%%%%%%%%%%%%%%%%%%%%%%%%%%%%%%%%%%%%%%%%%%%%%%%%%%%%%%%%%%%
In Fig.\ref{fig2}, we display the theoretical prediction of $Br(Z\rightarrow \nu\nu)$ versus Dirac mass parameter $M_D^W$ in MRSSM. The solid line and the dot line stand for the result calculated with parameter setup BMP1 with normal ordering and inverse ordering respectively. The dash dot line and the dash line stand for the result calculated with parameter setup BMP2 with normal ordering and inverse ordering respectively. Both predictions are compatible with the experimental measurement at $2\sigma$ level. Similar to the case in Fig.\ref{fig1}, the prediction of $Br(Z\rightarrow \nu\nu)$ with BMP1 is a little lower than that with BMP2 for normal ordering. However, the case is opposite for inverse ordering. For normal ordering, there is a peak around $M_D^W=$400 GeV due to the mixing among the susy particles. For BMP1, at range [100,300] GeV, [500,1000] GeV and around 400 GeV, the prediction of $Br(Z\rightarrow \nu\nu)$ exceeds the SM expectation at $1\sigma$ level. The allowed range is narrowed at 300 GeV $<M_D^W<$ 400 GeV and 400 GeV $<M_D^W<$ 500 GeV. For BMP2, the allowed range is narrowed at 200 GeV $<M_D^W<$ 400 GeV and 400 GeV $<M_D^W<$ 600 GeV. For inverse ordering, the prediction of $Br(Z\rightarrow \nu\nu)$ decreases slowly as $M_D^W$ increases. There is also a deviation between the predictions in MRSSM and the experimental measurement at $1\sigma$ level.

%%%%%%%%%%%%%%%%%%%%%%%%%%%%%%%%%%%%%%%%%%%%%%%%%%%%%%%%%%%%%%%%%%%
\begin{figure*}
\centering
\includegraphics[width=4.0in]{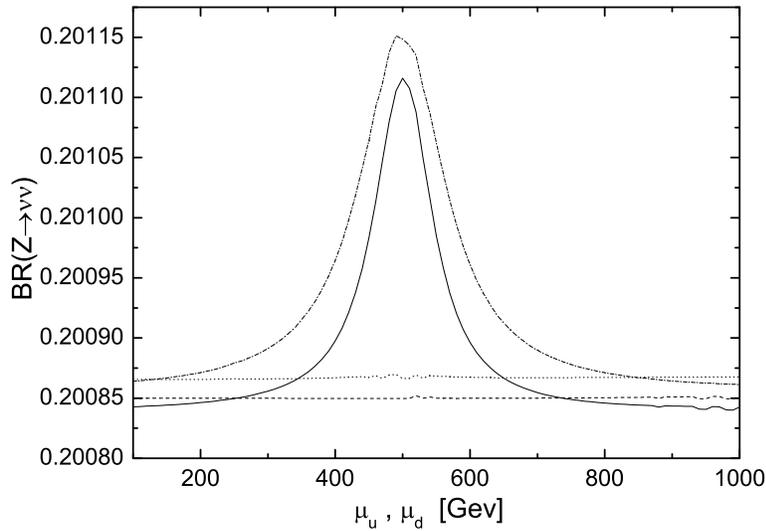}
\caption[]{The $Br(Z\rightarrow \nu\nu)$ vary as a function of $\mu_u(\mu_d)$ in MRSSM, where the solid line (BMP1) and the dash dot line (BMP2) correspond to normal neutrino mass ordering, the dot line (BMP1) and the dash line (BMP2) correspond to inverse neutrino mass ordering.   }
\label{fig3}
\end{figure*}
%%%%%%%%%%%%%%%%%%%%%%%%%%%%%%%%%%%%%%%%%%%%%%%%%%%%%%%%%%%%%%%%%%
%%%%%%%%%%%%%%%%%%%%%%%%%%%%%%%%%%%%%%%%%%%%%%%%%%%%%%%%%%%%%%%%%%%%
\begin{figure*}
\centering
\includegraphics[width=3.9in]{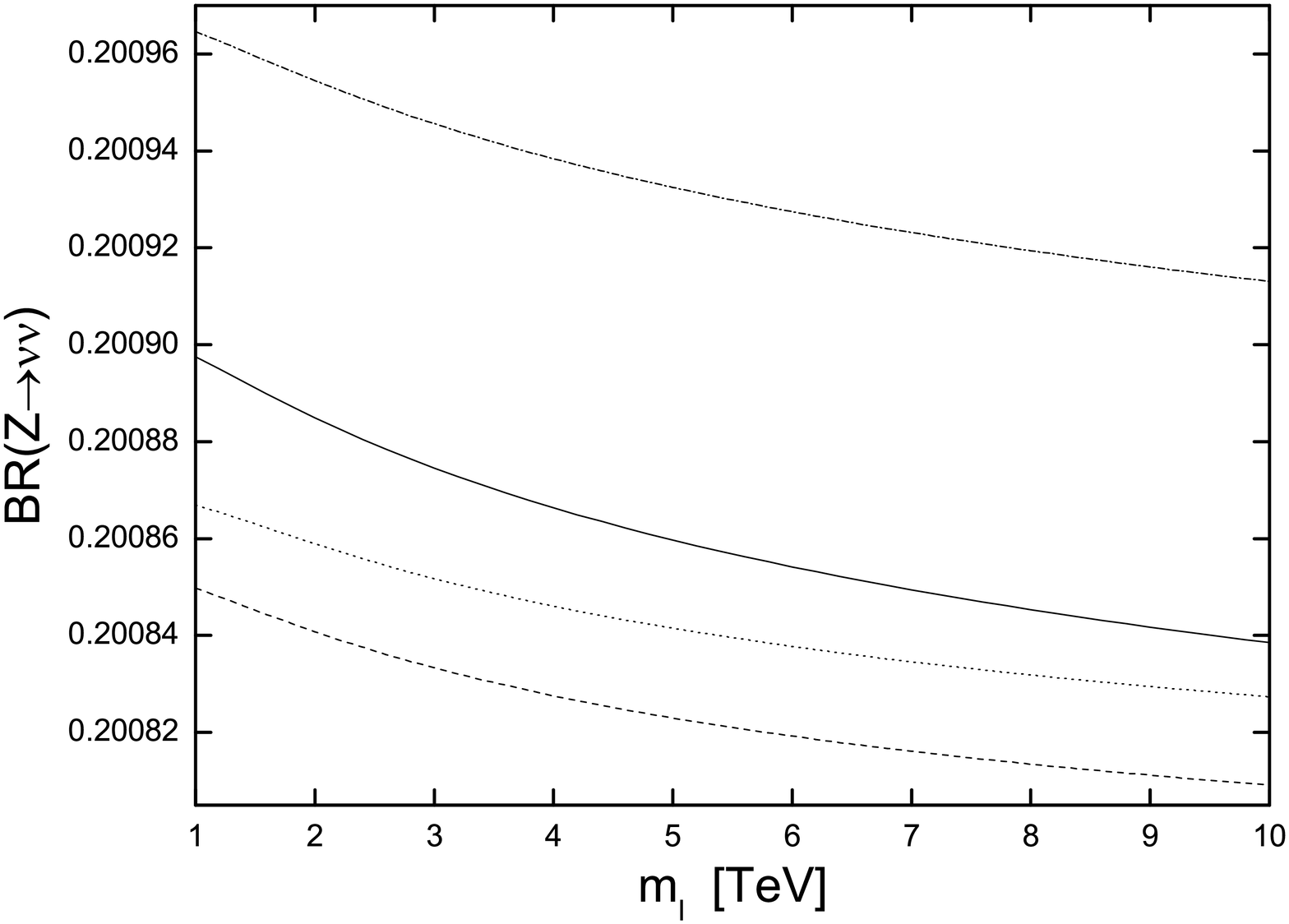}
\caption[]{The $Br(Z\rightarrow \nu\nu)$ vary as a function of $m_l$ in MRSSM, where the solid line (BMP1) and the dash dot line (BMP2) correspond to normal neutrino mass ordering, the dot line (BMP1) and the dash line (BMP2) correspond to inverse neutrino mass ordering.}
\label{fig4}
\end{figure*}
%%%%%%%%%%%%%%%%%%%%%%%%%%%%%%%%%%%%%%%%%%%%%%%%%%%%%%%%%%%%%%%%%%%

In Fig.\ref{fig3}, we display the theoretical prediction of $Br(Z\rightarrow \nu\nu)$ versus higgsino mass parameter $\mu_u(\mu_d)$ in MRSSM. The solid line and the dot line stand for the result calculated with parameter setup BMP1 with normal ordering and inverse ordering respectively. The dash dot line and the dash line stand for the result calculated with parameter setup BMP2 with normal ordering and inverse ordering respectively. Both predictions are compatible with the experimental measurement at $2\sigma$ level. Similar to the case in Fig.\ref{fig2}, the prediction of $Br(Z\rightarrow \nu\nu)$ with BMP1 is lower than BMP2 with normal ordering and higher than BMP2 with inverse ordering. For normal ordering, due to the mixing among the susy particles, there is also a peak around $\mu_u=\mu_d=$500 GeV. The values at the peak exceeds the SM expectation and the allowed range is narrowed at 400 GeV $<\mu_u(\mu_d)<$ 600 GeV for BMP1 and 300 GeV $<\mu_u(\mu_d)<$ 700 GeV for BMP2 at $1\sigma$ level. For inverse ordering, the prediction of $Br(Z\rightarrow \nu\nu)$ is almost invariant as $\mu_u(\mu_d)$ vary. A deviation between the predictions in MRSSM and the experimental measurement at $1\sigma$ level is displayed.

In Fig.\ref{fig4}, we display the theoretical prediction of $Br(Z\rightarrow \nu\nu)$ versus diagonal entries of slepton mass matrices $m_l(m_r)$ in MRSSM. The solid line and the dot line stand for the result calculated with parameter setup BMP1 with normal ordering and inverse ordering respectively. The dash dot line and the dash line stand for the result calculated with parameter setup BMP2 with normal ordering and inverse ordering respectively. Both predictions are compatible with the experimental measurement at $2\sigma$ level and decrease as $m_l(m_r)$ increases. Similar to the case in Fig.\ref{fig2}, the prediction of $Br(Z\rightarrow \nu\nu)$ with BMP1 is lower than BMP2 with normal ordering and higher than BMP2 with inverse ordering. For inverse ordering, the predictions are compatible with the SM expectation at $2\sigma$ level, but not compatible with the SM expectation at $1\sigma$ level.

\section{Conclusions}
\label{sec4}
We investigate the $Z\rightarrow \nu\nu$ decay in the framework of Minimal R-symmetric
Supersymmetric Standard Model (MRSSM) as a function of model parameters with normal and inverse neutrino mass ordering. Our analysis reveals that the theoretical prediction of $Br(Z\rightarrow \nu\nu)$ is compatible with the SM expectation and experimental measurement at $2\sigma$ level. At $1\sigma$ level, the theoretical prediction of $Br(Z\rightarrow \nu\nu)$ with normal ordering is more compatible with the SM expectation, and exceeds the SM expectation for inverse ordering. For inverse ordering, the $Br(Z\rightarrow \nu\nu)$ is between the SM expectation and experimental measurement. Finally, for a careful consideration, more precise measurement of $Br(Z\rightarrow invisible)$ in experiment is in need.

\section*{Acknowledgements}
The work has been supported by the National Natural Science Foundation of China (NNSFC) with Grants No.11747064, No.11805140 and No.11705045, the Scientific and Technological Innovation Programs of Higher Education Institutions in Shanxi with Grant No. 2017113, the Foundation of Department of Education of Liaoning province with Grant No. 2016TSPY10 and the Youth Foundation of the University of Science and Technology Liaoning with Grant No. 2016QN11.

\end{document}